\documentclass[10pt,a4paper,onecolumn,aps,nofootinbib,showpacs]{revtex4}

\usepackage[T1]{fontenc}
\usepackage[cp1250]{inputenc}

\usepackage{graphicx}
\usepackage{placeins}
\usepackage{dsfont}
\usepackage{amsmath,amssymb,graphics}

\usepackage{color}
\usepackage{bm}

\begin{document}

\title{SUPER-EXTREME EVENT'S INFLUENCE ON A WEIERSTRASS-MANDELBROT CONTINUOUS-TIME RANDOM WALK} 
\author{Tomasz Gubiec} 
\email{Tomasz.Gubiec@fuw.edu.pl}
\affiliation{Institute of Experimental Physics, Faculty of Physics \\ University of Warsaw, 
Smyczkowa Str. 5/7, PL-02678 Warsaw, Poland}
\author{Tomasz R. Werner}
\email{Tomasz.Werner@fuw.edu.pl}
\affiliation{Institute of Theoretical Physics, Faculty of Physics \\ University of Warsaw, 
Ho\.za 69, PL-00681 Warsaw, Poland}
\author{Ryszard Kutner}
\email{Ryszard.Kutner@fuw.edu.pl (correspondence_author)}
\affiliation{Institute of Experimental Physics, Faculty of Physics \\ University of Warsaw, 
Smyczkowa Str. 5/7, PL-02678 Warsaw, Poland} 
\author{Didier Sornette}
\email{dsornette@ethz.ch}
\affiliation{ETH-Z\"urich \\ 
Department of Management, Technology and Economics \\ Kreuzplaz 5, CH-8032 Z\"urich, Switzerland}

\begin{abstract}
Two utmost cases of \emph{super-extreme event's} influence on the velocity autocorrelation function (VAF) were 
considered. The VAF itself was derived within the hierarchical Weierstrass-Mandelbrot Continuous-Time Random 
Walk (WM-CTRW) formalism, which is able to cover a broad spectrum of continuous-time random walks. Firstly, we 
studied a super-extreme event in a form of a sustained drift, whose duration time is much longer than that of any 
other event. Secondly, we considered a super-extreme event in the form of a shock with the size and velocity  
much larger than those corresponding to any other event. We found that the appearance of these super-extreme events 
substantially changes the results determined by extreme events (the so called "black swans") that are endogenous 
to the WM-CTRW process. For example, changes of the VAF in the latter case are in the form of some 
instability and distinctly differ from those caused in the former case. In each case these changes are quite 
different compared to the situation without super-extreme events suggesting the possibility to detect them in 
natural system if they occur. 
\end{abstract}


\maketitle

\section{Introduction} 

One of the most remarkably emerging observation within the natural and socio-economical sciences is that  
the empirical data which they supply are frequently punctuated by rare extreme events or \emph{black swans} which can 
play a dominant role. This observation is usually quantified by power-laws or heavy-tailed probability distributions 
of event sizes (cf. \cite{DS0, AJK,BH,BH1,HA,AH,GB,MK,SZF,KPS,MS,MDS,DS} and references therein). However, as it was
pointed out in \cite{DS}, there is an empirical evidence that something important beyond power-laws does exist. 
In this context, the concept of \emph{super-extreme events}, \emph{outliers} or \emph{dragon-kings} was introduced.

By the super-extreme event, outlier or dragon-king\footnote{The poetic term "dragon-king" stresses that: (i) we deal 
with exceptional event which is a completely different kind of "animal" in comparison with the usual ones concerning  
the rest of events, (ii) it emphasizes the importance of this event being outside the power-law like an absolute 
monarchy standing above the law or as the wealth of a monarch owning a finite part of the whole country,
beyond the Pareto distribution of its citizen's wealths \cite{LS}.} we mean an event of the values that are abnormally different than values 
of other events in a random sample taken from a given population \cite{ESH}. It is therefore an anomaly, an event 
usually removed in order to obtain reliable statistical estimations. The term "outlier" emphasizes the spurious nature 
of these anomalous events, suggesting discarding them as errors or misleading monsters. In contrast, the term 
"dragon-king" emphasizes their relevance in the dynamics and the importance of keeping them to be able to understand
the generating process. The super-extreme 
events are statistically complementary to extreme events, as documented in \cite{DS}. Moreover, 
the idea that dragon-kings are often associated with the occurence of a catastrophe, 
a phase transition, and bifurcation as well as with a tipping point, whose emergent organization produces visible 
precursors was also developed there.
 
The main goal of the present work is to analytically demonstrate and numerically simulate two utmost cases of the 
influence of dragon-kings on the velocity autocorrelation function (VAF) of a random walker. Herein, we studied VAF 
in the frame of the Weierstrass-Mandelbrot Continuous-Time Random Walk (WM-CTRW) formalism developed in 
\cite{KR,KS,KutS}. This model is a hierarchical version of the canonical Continuous-Time Random Walk (CTRW) formalism 
\cite{PS,HK,KK} in which the hierarchical spatio-temporal waiting-time distribution (WTD) 
was assumed as its basic quantity (cf. Section \ref{section:WCTRW} in this work as well as Equation (20) in \cite{KS}). 

Note that the WM-CTRW formalism is sufficiently generic and flexible. It is able to cover various types of diffusion,
i.e. from normal diffusion, through the superdiffusion (e.g. the persistent fractional Brownian motion, fBm, 
\cite{TT}), and the ballistic one, to the L\'evy walk. Remarkably, the extreme events are contained in the 
spatio-temporal structure of time series obtained within the WM-CTRW formalism by stochastic simulation. This is 
explained in details in Section \ref{section:Defprobl}. The extreme events moderate the relaxation of the system to 
equilibrium (or partial equilibrium), e.g. by changing the relaxation from exponential to power-law. Furthermore, 
we used the WM-CTRW formalism
because we (superficially) verified that the hierarchical WTD described quite well its empirical counterparts 
obtained for continuous quotation of the exchange rates on a currency exchange market as well as of the share price 
trading on a stock exchange.

The present paper is organized in the following manner. First (in Section \ref{section:Defprobl}), our problem is 
defined along with definitions of the most relevant quantities used in our study. Then (Section \ref{section:RWCTRW}),
the WM-CTRW formalism is defined and, next, a superdiffusion phase is discussed. Subsequently, Section 
\ref{section:CdCDt} presents derivation of VAF including a dragon-king event. The comparison of predictions of our
theoretical formulae for the VAF with the results of the simulations is contained in Section \ref{section:cws}. 
Finally, summary and concluding remarks are presented in Section \ref{section:sumconclud}. 

\section{Definition of the problem}\label{section:Defprobl}

In the present work we consider, as an example, a superdiffusion case within the WM-CTRW formalism, where  
so-called weak ergodicity is obeyed \cite{MB,BB,BeBa}; that is, the mean value of the waiting-times between turning 
points of a random walk trajectory is finite. 
As the probability of appearance of a dragon-king is extremely small (that is, the waiting time for its appearance is
too long for all practical purposes), it was produced after the simulation and next put "manually" somewhere inside the 
series. In that sense, the dragon-king's appearance can be considered as an exogenous event. 

\begin{figure}[htb]
\begin{center}
\includegraphics[width=140mm,angle=0,clip]{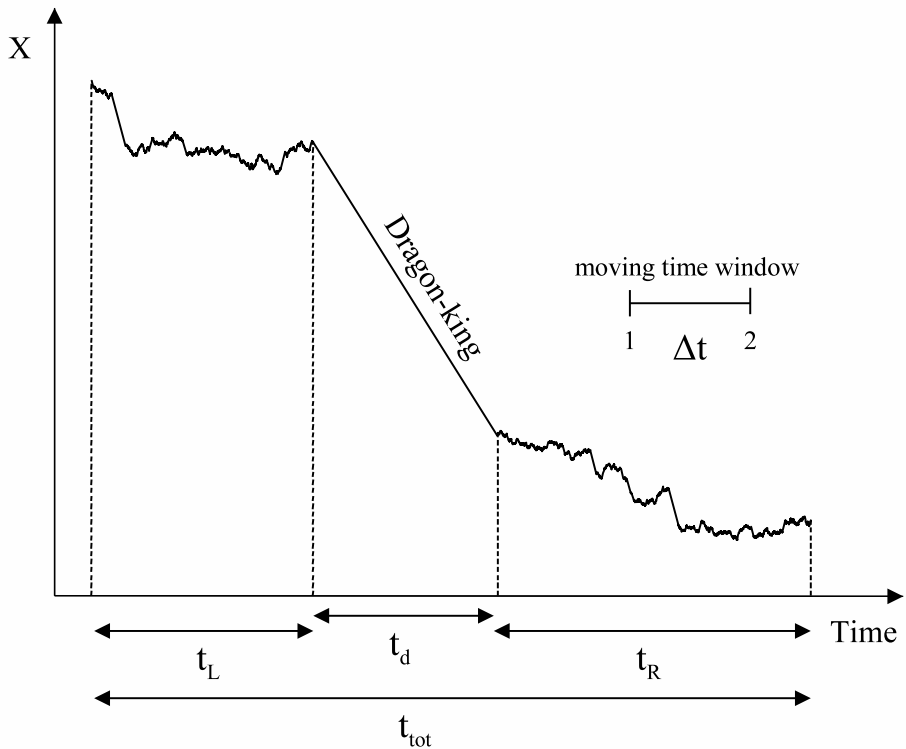}
\caption{Schematic time series for the total time 
$t_{tot}$ containing the dragon-king event (represented by the longest sloping straight line) having constant velocity 
$v_d$ and duration time $t_d$. The time series before and after the occurrence of the dragon-king
have duration times $t_L$ and $t_R$, respectively. Obviously, $t_{tot}=t_L+t_d+t_R$.}
\label{figure:dragon_1}
\end{center}
\end{figure}

\begin{figure}[htb]
\vspace{-1.0cm}
\begin{center}
\includegraphics[width=140mm,angle=0,clip]{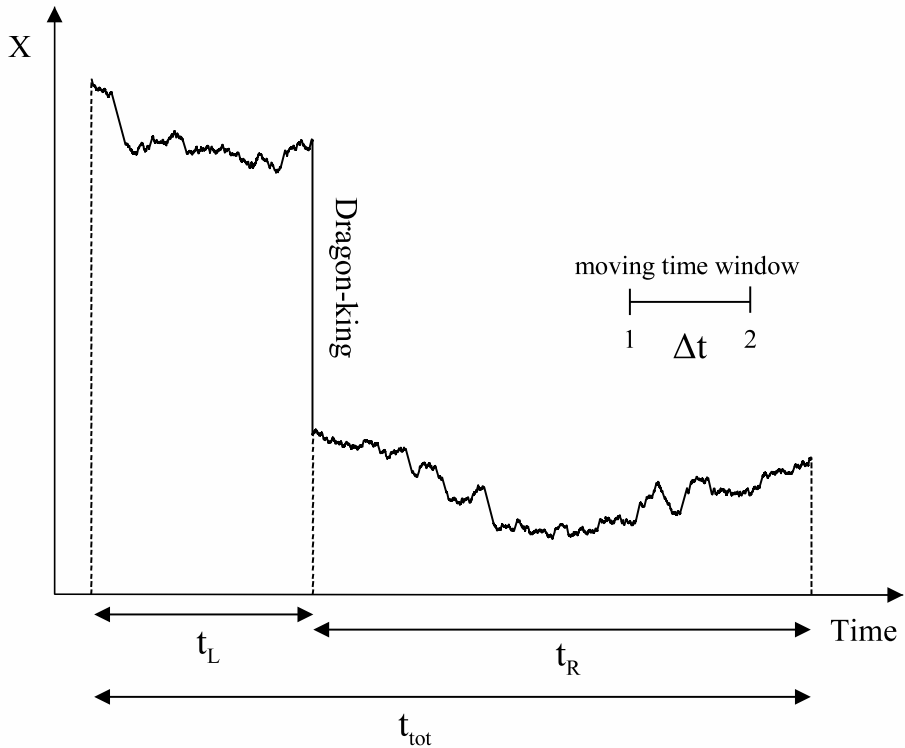}
\caption{Schematic time series containing a dragon-king in the form of a shock (represented by the longest vertical 
straight line). The time series before and after the dragon-king have duration times $t_L$ and $t_R$, 
respectively while the duration time of the shock is $t_d=dt$ being so short that it cannot be visualized in the plot 
($t_{tot}=t_L+dt+t_R$).}
\label{figure:Shock}
\end{center}
\end{figure}

We answer the question of how much the stationary velocity autocorrelation function, 
$C(\Delta t)$, derived from a given time series is changed when this time series is suddenly punctuated by 
a single-step super-extreme event. We consider two types of super-extreme events: 
\begin{itemize}
\item[(i)] the long-drawn event which has super-extremely long duration time $t_d$ (cf. Figure \ref{figure:dragon_1}) 
and 
\item[(ii)] the shock, or sudden jump, of a random variable $X$, which has super-extreme velocity $v_d$ 
(cf. Figure \ref{figure:Shock}). 
\end{itemize}
The background of the corresponding dragon-kings' definition is again presented in Figures \ref{figure:dragon_1} and 
\ref{figure:Shock}. The resulting VAF, involving a dragon-king, is denoted below by $C_d(\Delta t)$. 

The VAF is defined as 
\begin{eqnarray}
\mbox{VAF}(\Delta t)&=&\left<v(t')\, v(t'+\Delta t)\right>-\left<v(t')\right>\left<v(t'+\Delta t)\right> \nonumber \\
&=&\left<v_1\, v_2\right>(\Delta t)-\left<v_1\right>\left<v_2\right>=
\left\{
\begin{array}{cc}
C(\Delta t), & \mbox{in the absence of a dragon-king,} \\
C_d(\Delta t), & \mbox{in the presence of a dragon-king,}
\end{array}
\label{rown:vaf1}
\right.
\end{eqnarray}
where $\left<\ldots \right>$ means the moving average (or averaging over current time $t'\leq t_{tot}-\Delta t$) in a 
given time-window $\Delta t$ (cf. Figure \ref{figure:dragon_1} and \ref{figure:Shock} and considerations in Section 
\ref{section:CdCDt}) and $v(t')\stackrel{\rm def.}{=}[X(t')-X(t'\, - dt)]/dt$, where the time-discretization step is
$dt\leq \Delta t,\, t'$. The velocities $v_1$ and $v_2$ are defined here at the beginning  and at the 
end of the time-window $\Delta t$, respectively. 
   
Both quantities, $C(\Delta t)$ and $C_d(\Delta t)$, are studied analytically (in Sections \ref{section:RWCTRW} and 
\ref{section:CdCDt}) and by numerical simulations (in Section \ref{section:cws}) because our task is to find  
relations between $C(\Delta t)$ and $C_d(\Delta t)$ and verify them by simulations for cases (i) 
and (ii) mentioned above.

\section{Weierstrass-Mandelbrot Continuous-Time Random Walk}\label{section:RWCTRW}

The task of this section is to briefly sketch elements of the WM-CTRW formalism that are useful for our analysis. 
An explicit form of the VAF, $C(\Delta t)$, in the absence of a dragon-king was described in details in our earlier 
works \cite{KS,KutS}.

\subsection{Definition of the WM-CTRW formalism}\label{section:WCTRW}

The Weierstrass-Mandelbrot Continuous-Time Random Walk is defined by the hierarchical spatio-temporal waiting-time 
distribution (WTD).
This WTD\footnote{The complete definition of the WM-CTRW model additionally requires a special treatment of the first 
step of the random walker \cite{KS}. However, it is irrelevant when moving-average is performed. Therefore, we 
do not consider this special treatment in this work.} is given by the following weighted hierarchical series
\begin{eqnarray}
\psi (x,\, t)=\sum_{j=0}^{\infty }w(j)\psi _j(x,\, t),
\label{rown:psit}
\end{eqnarray}
where $x$ is the walker single-step spatial displacement passed (with constant velocity) within the time interval $t$ 
and the weight $w(j)$ is given by the probability distribution 
\begin{eqnarray}
w(j)=\frac{1}{N^j}\left(1-\frac{1}{N}\right),\; N>1,\; \; j=0,1,2,\, \ldots \, .
\label{rown:wj}
\end{eqnarray}
This weight can be interpreted as the probability of exactly $j$ consecutive successes in some Bernoulli series,
where $1/N$ is the probability of a single success and the conditional single-level WTD is assumed in the factorized 
form of two different single-variable distributions $f$ and $h$
\begin{eqnarray}
\psi _j(x,\, t)=\frac{1}{v_0v^jt}f\left(\frac{\mid x\mid }{v_0v^jt}\right)
\frac{1}{\tau _0\tau ^j}h\left(\frac{t}{\tau _0\tau ^j}\right).
\label{rown:psijt}
\end{eqnarray}
In Expression (\ref{rown:psijt}), we used a simple representation of a random walk (but \underline{not} random jumps 
or flights). That is, conditional temporal and spatial probability distributions we assumed are of the forms
\begin{eqnarray}
f\left(\frac{\mid x\mid }{v_0v^jt}\right)=\frac{1}{2}\, \delta \left(\frac{\mid x\mid }{v_0v^jt}-1\right)
\label{rown:htj}
\end{eqnarray}
and
\begin{eqnarray}
h\left(\frac{t}{\tau _0\tau ^j}\right)=\exp\left(-\frac{t}{\tau _0\tau ^j}\right),
\label{rown:fXj}
\end{eqnarray}
respectively\footnote{Obviously, detailed forms of the scaling functions $f$ and $h$ are less important for macroscopic 
displacement and asymptotic long time, respectively. Here, we used their simple explicit forms to make our 
calculations easier.}. The mean duration time, $\tau _0\tau ^j$, of the random walker single step and its velocity, 
$v_0v^j$, are both associated with the $j^{th}$ level of the spatio-temporal hierarchy. This level is the same for 
temporal and spatial partial probability distributions, introducing the spatio-temporal coupling in  
(\ref{rown:psit}). Herein, we marked the calibration parameters by $\tau _0$ and $v_0$.

As it is seen from (\ref{rown:psit}), our WM-CTRW formalism belongs to the \emph{non-separable} category of 
CTRW, which makes the diffusion phase diagram quite rich \cite{KR,KS}. 

From (\ref{rown:psit}), it is easy to derive  useful temporal and spatial single-step moments 
\begin{eqnarray}
\left<t\right>=
\int_0^{\infty }dt\, t\int_{-\infty }^{\infty }dx\, \psi (x,\, t)=
\tau_0\, \left\{
\begin{array}{cc}
\frac{1-1/N}{1-\tau /N}, & \mbox{for $\alpha >1$,} \\
\infty , & \mbox{for $\alpha <1$}
\label{rown:tttt}
\end{array}
\right.
\end{eqnarray}
where $\alpha =\ln N/\ln \tau $ and
\begin{eqnarray}
\left<x^2\right>=
\int_{-\infty }^{\infty }dx\, x^2\int_0^{\infty }dt\, \psi (x, \, t)=
(b_0)^2\, \left\{
\begin{array}{cc}
2\, \frac{1-1/N}{1-b^2/N}, & \mbox{for $\beta >2$,} \\
\infty , & \mbox{for $\beta <2$}
\label{rown:x22}
\end{array}
\right.
\end{eqnarray}
where $b_0=v_0\tau _0,\; b=v\tau $ and $\beta =\ln N/\ln b$.

Notably, $\alpha $ and $\beta $ are parameters which control the different phases of diffusion. It was proved in 
\cite{KS} that the (total) fractional diffusion exponent $\eta $, which governs the asymptotic time-dependence 
of the walker multi-step mean-square displacement (cf. Equation (11) in \cite{KS}), 
\begin{eqnarray} 
\left<X^2(\Delta t)\right>\approx \frac{2\, D_{st}}{\Gamma (\eta +1)}\, (\Delta t)^{\eta },
\label{rown:msd}
\end{eqnarray}
itself depends on $\alpha $ and $\beta $ (cf. Table 1 and diffusion phase diagram shown in Figure 1 in \cite{KS}). 
We term these parameters \emph{temporal} and \emph{spatial diffusion exponents}\footnote{It can be also proved that 
there exists a simple exponent, which is a function of the \emph{partial diffusion exponents} $\alpha $ and $\beta $, 
that governs the time-dependence of the walker single-step mean-square displacement, 
$\left<[x(t)]^2\right>=
\int_{-\infty }^{\infty }dx\, x^2\psi (x,\, t)$, for asymptotic time interval 
$t$.}, respectively. The quantity $D_{st}$ is the so-called \emph{fractional diffusion coefficient} emphasizing its 
association with the stationarized random walk 
\cite{KS}. It depends on partial (temporal and spatial) diffusion exponents (see also  
Table 1 in \cite{KS}). Notably, the total time is $t_{tot}=\sum t$ and displacement 
of the walker is $X(t_{tot})=\sum x$ (where $\sum $ means the sum over successive steps of the walker).

Expressions (\ref{rown:psit}) - (\ref{rown:fXj}) enable simulation of a random walk trajectory (schematically shown in
Figures \ref{figure:dragon_1} and \ref{figure:Shock}) in continuous time. This is because they define the 
corresponding stochastic dynamics (considered in Section \ref{section:Wsda}). 

\subsection{Superdiffusion phase}\label{section:Ed}

The WM-CTRW defines a stationary stochastic process valid only for the case where the mean waiting-time, 
$\left<t\right>$, is finite, i.e. for the case where the temporal exponent $\alpha >1$ (cf. Equation (\ref{rown:tttt}) 
as well as Equations (4) and (5) in \cite{KS}). Under such conditions, the diffusion exponent $\eta =2H$, where 
$0<H\le 1$, is the well known Hurst exponent \cite{TT}.
All our analytical calculations are confined to the case where the (multi-step) mean-square displacement is finite 
for a finite time and superlinearly increases with time for asymptotically long time. That is, we are confined to the 
superdiffusion phase where $\eta >1$. This regime is of interest because dragon-kings could be argued to be least 
relevant in such phase. Our goal is to demonstrate that dragon-kings make a significant impact even in such 
a superlinear phase. 

The superdiffusion phase is restricted by the following inequalities
\begin{eqnarray}
\frac{1}{2}<\frac{1}{\beta }<\frac{1}{2}+\frac{1}{2\alpha }.
\end{eqnarray}

For the superdiffusion phase, we can easily derive the VAF, in the absence of a dragon-king, in the form (cf. Equation 
(13) in \cite{KS})
%{\color{blue}
\begin{eqnarray}
C(\Delta t) = \frac{2\, D_{st}}{\Gamma (\eta -1)}\frac{1}{\Delta t^{2-\eta }},
\label{rown:DXtst}
\end{eqnarray}
where the fractional diffusion coefficient is given by
\begin{eqnarray}
D_{st}=\frac{1-\frac{\tau }{N}}{\ln N}
\frac{\pi \alpha }{\sin\left(2\pi \alpha \left(\frac{1}{\beta }-\frac{1}{2}\right)\right)}
\label{rown:Dst}
\end{eqnarray}
and the fractional diffusion exponent is
\begin{eqnarray}
\eta =1+2\alpha \left(\frac{1}{\beta }-\frac{1}{2}\right).
\label{rown:etast}
\end{eqnarray} 

As the fractional diffusion exponent $\eta <2$, the velocity autocorrelation function, $C(\Delta t)$, given by 
Expression (\ref{rown:DXtst}) vanishes for extremely long $\Delta t$. In Section \ref{section:CdCDt} we prove that 
the presence of a dragon-king in the time series can lead to violation of this property. 

\section{Derivation of formulae for $C_d(\Delta t)$}\label{section:CdCDt}

In this section we derive relations of two different types between the estimator of $C_d(\Delta t)$ and the
estimators of some VAFs concerning time series in absence of any dragon-king. That is, in Section \ref{section:Gc} 
we consider case (i) while in Section \ref{section:shock} case (ii), where both were already defined in Section 
\ref{section:Defprobl}. 

\subsection{$C_d(\Delta t)$ for the case of sustained dragon-king}\label{section:Gc}

In this section we assume that 
\begin{itemize}
\item[(1)] $dt\le \Delta t \le \Delta t_{MAX}\ll t_L,\, t_R$ and 
\item[(2)] $\Delta t_{MAX}\le t_d$, where $\Delta t_{MAX}$ is a maximal value of $\Delta t$. 
\end{itemize}
Importantly, assumption (2) is violated for case (ii) defined in Section \ref{section:Defprobl} above and 
considered below in Section \ref{section:shock}.

We divide the expected value of the product of velocities, $\left<v_1\, v_2\right>(\Delta t)$, present in Formula 
(\ref{rown:vaf1}), into five weighted different components. They are estimated by the following terms
\begin{eqnarray}
\left<v_1\, v_2\right>(\Delta t)=\sum_{m=1}^{M=5}\left<v_1\, v_2\right>_m(\Delta t)\, w_m.
\label{rown:v1v25terms}
\end{eqnarray}
It is straightforward to derive each component $\left<v_1\, v_2\right>_m(\Delta t)$ separately by using the proper 
moving-average estimator. The corresponding weight $w_m$ is easy to obtain.
\begin{itemize}
\item[(1)] The first component (for $m=1$) is defined for the case, where both velocities $v_1$ and $v_2$ 
are placed before the dragon-king position. This component, together with the corresponding weight are, 
as follows
\begin{eqnarray}
\left<v_1\, v_2\right>_1(\Delta t)\stackrel{\rm def.}{=}
\frac{1}{t_L-\Delta t}\sum_{t'=dt}^{t_L-\Delta t}v(t')\, v(t'+\Delta t), \; \; \; \; w_1\stackrel{\rm def.}{=}
\frac{t_L-\Delta t}{t_{tot}-\Delta t}
\label{rown:v1v21}
\end{eqnarray}
and relates to a random walk in the absence of any dragon-king. More precisely, this component represents the 
situation where velocity $v_1$ is placed inside the time interval $[0,t_L-\Delta t]$ while velocity $v_2$ can be 
placed both inside this time interval or at its right border $t_L-\Delta t$. 
\item[(2)] For the second component ($m=2$), velocity $v_1$ is placed before the position of the 
dragon-king while velocity $v_2$ equals the velocity of the dragon-king $v_d$. The second component and the 
corresponding weight takes the form
\begin{eqnarray}
\left<v_1\, v_2\right>_2(\Delta t)\stackrel{\rm def.}{=}
\frac{1}{\Delta t}\sum_{t'=t_L-\Delta t+dt}^{t_L}v(t')\, v_d= \left<v\right>_Lv_d,\; \; \; \; 
w_2\stackrel{\rm def.}{=}\frac{\Delta t}{t_{tot}-\Delta t},
\label{rown:v1v22}
\end{eqnarray}
where $v_d$ is the velocity of the dragon-king (or the slope of the longest straight line shown in Figure 
\ref{figure:dragon_1}). This term describes the first cross situation where velocity $v_1$ is located inside the time 
interval $[t_L-\Delta t,t_L]$ or at its right border while velocity $v_2$ is placed inside the time-interval 
$[t_L,t_L+\Delta t]$ or at its right border. Obviously, $v_d$ is the constant velocity of the dragon-king.
\item[(3)] This case is particularly simple as both velocities $v_1$ and $v_2$ are equal to the dragon-king's velocity 
$v_d$. The third component ($m=3$) is the simplest one and together with the corresponding weight assume the 
forms
\begin{eqnarray}
\left<v_1\, v_2\right>_3(\Delta t)\stackrel{\rm def.}{=}v_d^2,\; \; \; \; w_3\stackrel{\rm def.}{=}
\frac{t_d-\Delta t}{t_{tot}-\Delta t},
\label{rown:v1v23}
\end{eqnarray}
and corresponds to the case where both velocities $v_1$ and $v_2$ are placed inside the dragon-king's time interval 
$[t_L,t_L+t_d]$ or velocity $v_2$ can be also located at its right border. 
\item[(4)] Herein, velocity $v_1$ equals the dragon-king's velocity $v_d$ while velocity $v_2$ is placed after the 
position of the dragon-king. The fourth component ($m=4$) and the weight are
\begin{eqnarray}
\left<v_1\, v_2\right>_4(\Delta t)\stackrel{\rm def.}{=}
\frac{1}{\Delta t}\sum_{t'=t_L+t_d-\Delta t+dt}^{t_L+t_d}v_d\, v(t')= v_d\, \left<v\right>_R,\; \; \; \;
w_4\stackrel{\rm def.}{=}\frac{\Delta t}{t_{tot}-\Delta t},
\label{rown:v1v24}
\end{eqnarray}
corresponding to the second cross case. Precisely, for this case velocity $v_1$ is placed inside the 
dragon-king time interval $[t_L+t_d-\Delta t,t_L+t_d]$ or at its right border while velocity $v_2$ is placed inside 
the time interval $[t_L+t_d,t_L+t_d+\Delta t]$ or at its right border.  
\item[(5)] For the fifth component ($m=5$), positions of both velocity $v_1$ and $v_2$ are placed after the position 
of the dragon-king. The fifth component and the weight are, as follows
\begin{eqnarray}
\left<v_1\, v_2\right>_5(\Delta t)\stackrel{\rm def.}{=}\frac{1}{t_R-\Delta t}
\sum_{t'=t_{tot}-t_R+dt}^{t_{tot}-\Delta t}v(t')\, v(t'+\Delta t),\; \; \; \;
w_5\stackrel{\rm def.}{=}\frac{t_R-\Delta t}{t_{tot}-\Delta t},
\label{rown:v1v25}
\end{eqnarray}
being analogous to those of the first component and weight. More precisely, they are defined for the case where both 
velocities $v_1$ and $v_2$ are placed inside the time interval $[t_{tot}-t_R,t_{tot}]$ (without of the dragon-king) 
and velocity $v_2$ can be also counted at time $t_{tot}$. 
\end{itemize}

Subsequently, we can calculate the estimator, which already includes the full influence of the sustained dragon-king
\begin{eqnarray}
C_d(\Delta t)&=&\left<v_1\, v_2\right>(\Delta t)-\left<v_1\right>\, \left<v_2\right> \nonumber \\
%&=&\frac{t_L-\Delta t}{t-\Delta t}\left<v_1\, v_2\right>_1(\Delta t)+
%\frac{\Delta t}{t-\Delta t}\left<v\right>_L\, v_d
%+\frac{t_d-\Delta t}{t-\Delta t}\, v_d^2 \nonumber \\
%&+&\frac{\Delta t}{t-\Delta t}v_d\, \left<v\right>_R 
%+\frac{t_R-\Delta t}{t-\Delta t}\left<v_1\, v_2\right>_5(\Delta t) \nonumber \\
%&-&\frac{1}{(t-\Delta t)^2}\left[t_L\left<v\right>_L+t_dv_d+
%(t_R-\Delta t)\left<v\right>_R\right]\left[(t_L-\Delta t)\left<v\right>_L+t_dv_d+
%t_R\left<v\right>_R\right] \nonumber \\
%&=&\frac{t_L-\Delta t}{t-\Delta t}\left<v_1\, v_2\right>_1(\Delta t)+
%\frac{t_R-\Delta t}{t-\Delta t}\left<v_1\, v_2\right>_5(\Delta t)+
%\frac{\Delta t}{t-\Delta t}\left(\left<v\right>_L+\left<v\right>_R\right)\, v_d+
%\frac{t_d-\Delta t}{t-\Delta t}\, v_d^2 \nonumber \\
%&-&\frac{1}{(t-\Delta t)^2}\left[t_L\left<v\right>_L+(t_R-\Delta t)\left<v\right>_R+t_dv_d\right]
%\left[(t_L-\Delta t)\left<v\right>_L+t_R\left<v\right>_R+t_dv_d\right] \nonumber \\
%&-&\left[\frac{(t-t_d-\Delta t)\left<v\right>+t_dv_d}{t-\Delta t}\right]^2
&=&\frac{\gamma _L-\Delta t/t_{tot}}{1-\Delta t/t_{tot}}\left<v_1\, v_2\right>_1(\Delta t)+
\frac{\gamma _R-\Delta t/t_{tot}}{1-\Delta t/t_{tot}}\left<v_1\, v_2\right>_5(\Delta t)+
\frac{\Delta t/t_{tot}}{1-\Delta t/t_{tot}}\left(\left<v\right>_L+\left<v\right>_R\right)v_d
+\frac{\gamma _d-\Delta t/t_{tot}}{1-\Delta t/t_{tot}}v_d^2 \nonumber \\
&-&\frac{1}{(1-\Delta t/t_{tot})^2}\left[\gamma _L\left<v\right>_L+(\gamma _R-\Delta t/t_{tot})\left<v\right>_R+
\gamma _dv_d\right]\left[(\gamma _L-\Delta t/t_{tot})\left<v\right>_L+\gamma _R\left<v\right>_R+\gamma _dv_d\right]
\label{rown:covd}
\end{eqnarray}
by using definitions
\begin{eqnarray}
\left<v_1\right>\stackrel{\rm def.}{=}\frac{1}{t_{tot}-\Delta t}\left[\sum_{t'=dt}^{t_L}v(t')+t_dv_d+
\sum_{t'=t_{tot}-t_R+dt}^{t_{tot}-\Delta t}v(t')\right]=\frac{1}{t_{tot}-\Delta t}\left[t_L\left<v\right>_L+
(t_R-\Delta t)\left<v\right>_R+t_dv_d\right]  \nonumber \\
\left<v_2\right>\stackrel{\rm def.}{=}\frac{1}{t_{tot}-\Delta t}\left[\sum_{t'=\Delta t+dt}^{t_L}v(t')+t_dv_d+
\sum_{t'=t_{tot}-t_R+dt}^{t_{tot}}v(t')\right]=\frac{1}{t_{tot}-\Delta t}\left[(t_L-\Delta t)\left<v\right>_L+
t_R\left<v\right>_R+t_dv_d\right], \nonumber \\
\label{rown:definitv1v2}
\end{eqnarray}
where $\left<v\right>_L$ and $\left<v\right>_R$ are partial mean velocities defined by the random walk on the left 
and right hand side of the dragon-king, respectively (cf. Figure \ref{figure:dragon_1}). We set velocities 
$\left<v_l\right>_L$ and $\left<v_l\right>_R$ equal to zero as no drift is present in the system. Besides, we used 
dimensionless parameters 
$\gamma _L\stackrel{\rm def.}{=}t_L/t_{tot},\, 
\gamma _R\stackrel{\rm def.}{=}t_R/t_{tot}$ and $\gamma _d\stackrel{\rm def.}{=}t_d/t_{tot}$.  

Because drift is absent in the system, Equation (\ref{rown:covd}), taking into account Definitions 
(\ref{rown:definitv1v2}), assumes a simpler form
\begin{eqnarray}
C_d(\Delta t)&=&\frac{\gamma _L-\Delta t/t_{tot}}{1-\Delta t/t_{tot}}\left<v_1\, v_2\right>_1(\Delta t)+
\frac{\gamma _R-\Delta t/t_{tot}}{1-\Delta t/t_{tot}}\left<v_1\, v_2\right>_5(\Delta t) \nonumber \\
%\frac{\Delta t/t}{1-\Delta t/t}\left(\left<v\right>_L+\left<v\right>_R\right)v_d \nonumber \\
%&+&\frac{\gamma _d-\Delta t/t}{1-\Delta t/t}v_d^2-\left(\frac{\gamma _dv_d}{1-\Delta t/t}\right)^2.
&+&\left[\frac{\gamma _d}{1-\Delta t/t_{tot}}\left(1-\frac{\gamma _d}{1-\Delta t/t_{tot}}\right)
-\frac{\Delta t/t_{tot}}{1-\Delta t/t_{tot}}\right]v_d^2.
\label{rown:Cfingen}
\end{eqnarray}
Apparently, the quantity $C_d(\Delta t)$ depends (in general) not only on the relative variable $\Delta t/t_{tot}$ 
but also on the parameters $\gamma _L,\, \gamma _R$, and $\gamma _d$. Hence, this quantity depends on the position 
of the dragon-king inside a time series, i.e. this quantity is, in general, a non-stationary one. However, it does 
not depend on the sign of the dragon-king velocity. 
Moreover, the origin of the dragon-king is irrelevant. That is, our derived formula is sufficiently 
generic in the sense that it is valid not only for the WM-CTRW but for any random walk.  

For sufficiently wide time window $\Delta t_{MAX}$ which still obeys $\Delta t_{MAX}/t_{tot}\ll 1$, the quantity 
$C_d(\Delta t)$ simplifies into the asymptotic formula
\begin{eqnarray}
C_d(\Delta t)\approx \left[\frac{\gamma _d}{1-\Delta t/t_{tot}}\left(1-\frac{\gamma _d}{1-\Delta t/t_{tot}}\right)
-\frac{\Delta t/t_{tot}}{1-\Delta t/t_{tot}}\right]v_d^2.
\label{rown:asympt}
\end{eqnarray}
As long as the dragon-king is present, $C_d(\Delta t)$ does not vanish, even though estimators 
$\left<v_1\, v_2\right>_1(\Delta t)$ and $\left<v_1\, v_2\right>_5(\Delta t)$ (which can strongly fluctuate) 
are decaying. 

Obviously, Formula (\ref{rown:asympt}) takes a simpler form
\begin{eqnarray}
C_d(\Delta t)\approx \gamma _d(1-\gamma _d)v_d^2
\label{rown:asymptsimpl}
\end{eqnarray}
if a strong but reasonable inequality 
\begin{eqnarray}
\frac{\Delta t_{MAX}}{t_{tot}}\ll min(\gamma _L,\, \gamma _R,\, \gamma _d \, (1-\gamma _d)) 
\label{rown:inequal}
\end{eqnarray}
is obeyed. 

\subsubsection{Important case of $\left<v_1\, v_2\right>_1(\Delta t)=\left<v_1\, v_2\right>_5(\Delta t)$} 

By assuming that random walks before and after the dragon-king appearance within the time series are statistically 
identical (although corresponding trajectories could be quite different) we can write, by neglecting unavoidable 
fluctuations, that $\left<v_1\, v_2\right>_1(\Delta t)=\left<v_1\, v_2\right>_5(\Delta t)=C(\Delta t)$. These 
equalities enable transforming Expression (\ref{rown:Cfingen}) into a stationary form
\begin{eqnarray}
C_d(\Delta t)=\frac{1-\gamma _d-2\Delta t/t_{tot}}{1-\Delta t/t_{tot}}\, C(\Delta t)+
\left[\frac{\gamma _d}{1-\Delta t/t_{tot}}\left(1-\frac{\gamma _d}{1-\Delta t/t_{tot}}\right)
-\frac{\Delta t/t_{tot}}{1-\Delta t/t_{tot}}\right]v_d^2,
\label{rown:CLCRC}
\end{eqnarray}
which is our reference formula. This stationary form is our achievement which enables several applications of Formula
(\ref{rown:CLCRC}).

The main difference between $C(\Delta t)$ and $C_d(\Delta t)$ is that the former  
asymptotically vanishes while the latter does not. This difference provides a tool 
which allows for distinguishing a power-law relaxation, controlled by rare extremes or black swans, from a decay 
controlled by the dragon-king. Indeed, we consider Expression (\ref{rown:CLCRC}) as a reference one, also relevant for 
more complex cases.

If Assumption (\ref{rown:inequal}) is valid, further simplification of Expression (\ref{rown:CLCRC}) can be made
\begin{eqnarray}
C_d(\Delta t)\approx (1-\gamma _d)\left[C(\Delta t)+\gamma _d\, v_d^2\right].
\label{rown:CLCRCsimpl}
\end{eqnarray}
This expression depends on two parameters $\gamma _d$ and $v_d$ fully characterizing the dragon-king. These 
parameters can be easily determined, e.g. from the initial and asymptotic nonvanishing values of $C_d(\Delta t)$. 
In fact, predictions of this simple formula is compared in Section \ref{section:cws} with corresponding results 
of simulation.

\subsection{$C_d(\Delta t)$ for the case of a shock (case (ii) of Section \ref{section:Defprobl})}\label{section:shock}

To consider the case of a shock within the time-series, we replace Assumtion (2), given in Section \ref{section:Gc}, 
by the following one
\begin{eqnarray}
t_d=dt.
\label{rown:assumpt3}
\end{eqnarray}
That is, the duration time of the dragon-king is assumed to be as short as possible and equals to the time 
discretization step $dt$. The Assumption (\ref{rown:assumpt3}) imposes a modification of components $m=2$ and $m=4$ in 
Expression (\ref{rown:v1v25terms}). Namely, components $m=2$ and $m=4$ together with their corresponding weights are 
replaced by expressions
\begin{eqnarray}
\left<v_1v_2\right>_2\stackrel{\rm def.}{=}v_L\, v_d,\; \; \; \; \; w_1=\frac{dt}{t_{tot}-\Delta t}, \hspace*{6.0cm}
\nonumber \\
\mbox{and} \hspace*{17.0cm}
\nonumber \\
\left<v_1v_2\right>_4\stackrel{\rm def.}{=}v_d\, v_R,\; \; \; \; \; w_2=\frac{dt}{t_{tot}-\Delta t}, \hspace*{6.0cm}
\nonumber 
\end{eqnarray}
respectively. Furthermore, instead of component $m=3$ we have 
\begin{eqnarray}
\left<v_1v_2\right>_3,\; \; \; \; \; w_3=\frac{\Delta t-dt}{t_{tot}-\Delta t},
\end{eqnarray}
where the current velocities $v_1=v_L$ and $v_2=v_R$ are now located before and after the position of the shock, 
respectively. Hence, a new relation for $C_d(\Delta t)$ has the form
\begin{eqnarray}
C_d(\Delta t)&=&\left[(t_L-\Delta t)\left<v_1\, v_2\right>_1(\Delta t)+dt\, v_Lv_d+
(\Delta t-dt)\left<v_1\, v_2\right>_3(\Delta t)+dt\, v_dv_R+(t_R-\Delta t)\left<v_1\, v_2\right>_5(\Delta t)\right]
\frac{1}{t_{tot}-\Delta t}\nonumber \\
&-&\left[t_L\left<v\right>_L+dt\, v_d+(t_R-\Delta t)\left<v\right>_R\right]
\left[(t_L-\Delta t)\left<v\right>_L+dt\, v_d+t_R\left<v\right>_R\right]\frac{1}{(t_{tot}-\Delta t)^2}.
\label{rown:dragonsh}
\end{eqnarray}

By analogy to case (i) of a sustained dragon-king, we set 
$\left<v_1\, v_2\right>_1(\Delta t)=\left<v_1\, v_2\right>_3(\Delta t)=\left<v_1\, v_2\right>_5(\Delta t)=C(\Delta t)$
that, in the case of no drift present in the system, simplifies (\ref{rown:dragonsh}) into the form 
\begin{eqnarray}
C_d(\Delta t)=\frac{t_{tot}-\Delta t-2dt}{t_{tot}-\Delta t}\, C(\Delta t)+\frac{(v_L+v_R)\, X_d}{t_{tot}-\Delta t}
-\left(\frac{X_d}{t_{tot}-\Delta t}\right)^2,
\label{rown:dragonsho}
\end{eqnarray}
where $X_d=dt\, v_d$ is the value of the shock. Note that $v_L$ and $v_R$ are velocities of the walker, which are 
separated by the time interval $2\Delta t$. From definition of the time step $dt$ we have $dt\ll t_{tot}-\Delta t$. 
Hence, Equation (\ref{rown:dragonsho}) again assumes a simpler form
\begin{eqnarray}
C_d(\Delta t)=C(\Delta t)+\frac{(v_L+v_R)\, X_d}{t_{tot}-\Delta t}
-\left(\frac{X_d}{t_{tot}-\Delta t}\right)^2,
\label{rown:dragonshoc}
\end{eqnarray}
which is our second basic formula. This formula is further modified in Section \ref{section:cws} to make  
$C_d(\Delta t)$ better suited for comparison with our results obtained by simulations.

\section{Algorithm and results}\label{section:cws}

\subsection{Stochastic dynamics}\label{section:Wsda}

The stochastic dynamics or algorithm simulating successive single-step displacements of the 
walker following a WM-CTRW consists of three stages. 
\begin{itemize}
\item[(i)] The drawing of the level index $j$ from distribution (\ref{rown:wj}) in each 
spatio-temporal step separately.
\item[(ii)] The calculation of the duration time $t$ of the single-step (or its elapsed time) from the
stochastic equation
\begin{eqnarray}
t=-\tau _0\tau ^j\, \ln (1-R),
\label{rown:dynstatt}
\end{eqnarray}
where $R\in [0,1[$ is a random number drawn from the random number generator (or from the uniform distribution 
confined to a unit interval); note that Equation (\ref{rown:dynstatt}) is equivalent to Equation (\ref{rown:fXj}) 
(after an application of well known method of the cumulative distribution function inversion).
\item[(iii)] The determination of the single-step displacement by Equation
\begin{eqnarray}
x(t)=\xi \, v_0\, v^jt,
\label{rown:dynstatX}
\end{eqnarray}
where stochastic variable $\xi $ is a dichotomic noise (i.e. $\xi =+1$ or $-1$ with equal probability $1/2$) and
$x(t)\stackrel{\rm def.}{=}X(t')-X(t'-t)$, where $t'$ is a current time (and not a time interval).
\end{itemize}

In this way construction of a single continuous in time trajectory of a random walk is possible. Obviously, 
the parameters $\tau _0, \tau , v_0$, and $v$ were fixed at the beginning of a whole simulation. Further in the 
text and in all our simulations we set the calibrating parameters $\tau _0=1$ and $v_0=1$. In fact, this trajectory 
is constructed in the frame of a non-stationarized version of a CTRW, where there is no special treatment of the 
initial step \cite{HK,KR,KS}. That is, the hierarchical spatio-temporal waiting-time distribution (\ref{rown:psit})
for the initial step is the same as for all other steps. Indeed, the stationarized  version of the CTRW, which 
differently treats the waiting-time distribution for the initial step, corresponds to the WM-CTRW formalism\footnote{It 
can be proved that the WM-CTRW formalism, even asymptotically, is more general than the fractional Brownian motion 
introduced by Mandelbrot and van Ness \cite{TT}. It is because the propagator derived within the WM-CTRW formalism 
can be non-Gaussian (for $1/\beta >1/\alpha $).} \cite{KS}. The VAF obtained theoretically within the WM-CTRW 
enables for comparison with the corresponding VAF obtained from simulations. This is possible because in 
simulations we deal with moving-averages, which by definition, average over the initial state, thus
supplying the required asymptotic stationary VAF.

The above given algorithm was used in Section \ref{section:comparis} to simulate the required basic random walk 
trajectory. The trajectory constructed in such a way is punctuated "manually" by the sustained dragon-king or 
by the shock dragon-king. However, preparation of the former dragon-king requires some explanation. 

\subsubsection{Preparation of sustained dragon-king}\label{section:sdk}

The stochastic dynamics defined by Equations (\ref{rown:dynstatt}) and (\ref{rown:dynstatX}) is controlled by three 
random variables: $j,\, R$, and $\xi $, while only the integer level index $j$, drawn from the distribution 
(\ref{rown:wj}), and random number $R$ are responsible for the size of a single step. For the case of the sustained 
dragon-king this dynamics is simplified by replacing Equation (\ref{rown:dynstatt}), which simulates 
an exponential distribution of interevent times (\ref{rown:fXj}), by the corresponding reduced expression 
\begin{eqnarray}
t=t_d=\tau _0\tau ^{j_d}.
\label{rown:sustaineddk}
\end{eqnarray}
That is, the size of the sustained super-extreme event is controlled only by a single random variable $j=j_d$ as 
are the velocity of the sustained dragon-king $v_d=v_0\, v^{j_d}$, its duration time $t_d=\tau _0\, \tau ^{j_d}$, 
and displacement $x_d=\xi \, b_0\, b^{j_d}$, where $b_0=v_0\, \tau _0$ and $b=v\, \tau $.  

We can say that the stochastic process defined by Equations (\ref{rown:sustaineddk}) and (\ref{rown:dynstatX}) 
is simplified. This process is a dicrete in time and hence in space, where 
\begin{eqnarray}
x(t)=X_d=\xi \, v_0\, \tau _0\, v^{j_d}\, \tau ^{j_d}=\xi \, b_0\, b^{j_d}.
\label{rown:dynstatbj}
\end{eqnarray}

If the index $j_d$ of a given dragon-king is fixed, the interevent times for this process cannot fluctuate. Its 
discrete-time step equals to the mean-time defining the exponential distribution (\ref{rown:dynstatt}) or 
(\ref{rown:fXj}). 

Apparently, we deal with two stochastic processes: (i) the first one which prepares the WM-CTRW trajectory or 
stochastic 
spatio-temporal hierarchy of events and (ii) the second process which generates the sustained dragon-king from the 
simplified, discrete in space and time hierarchical random walk. 
In fact, the latter process is used in Section \ref{section:stMW-CTRW} to illustrate the definition of black swan.

Right now, it is easy to separate in simulation the sustained dragon-king from the rest of events belonging to the
stochastic spatio-temporal hierarchy of events (cf. Section \ref{section:stMW-CTRW}). Namely, it is sufficient to 
choose the level index $j_d$ much larger than the maximal level index $j=j_{MAX}$ of the hierarchy defining 
the longest single-step displacement of any other event. The above described simplicity is the main reason for such 
a way of selection of a sustained dragon-king. 

\subsubsection{Hierarchical random walk and black swans}\label{section:stMW-CTRW}

It is decisive for our considerations that the ratio of successive weights
\begin{eqnarray}
\frac{w(j+1)}{w(j)}=\frac{1}{N},
\label{rown:wwjj}
\end{eqnarray}
is already $j$ independent. This means that steps defined by the level index $j$ are $N$ times more likely than those 
of the next step of larger order $j+1$. Therefore, one expects (on the average) that the walker will perform $N^j$ 
shorter steps before performing the next step of larger order. Hence, we can explain how extreme events or black swans 
control the hierarchical spatio-temporal structure of events in the frame of a simplified hierarchical random walk.

Here we consider, as a typical quantity, the mean-square displacement of the process 
\begin{eqnarray}
\left<X^2\right>(L)&=&\left<\left(\sum_{l=1}^L\, x_l\right)^2\right>=L \left<x^2\right>,
\label{rown:X2}
\end{eqnarray}
where $\left<\ldots \right>$ denotes an ensemble average, $L$ is the total number of the random walk steps, $x_l$ is 
a single-step displacement and $\left<x^2\right>$ is its mean-square value. In this derivation we neglected the 
off-diagonal term $\sum_{l\neq l'}^L\left<x_l\, x_{l'}\right>$ or crossed correlations between successive single-step 
displacements in comparison with the diagonal term $L \left<x^2\right>$. This is because of the process definition, 
which invokes independent draw of steps.  

\begin{figure}[htb]
\begin{center}
\includegraphics[width=140mm,angle=0,clip]{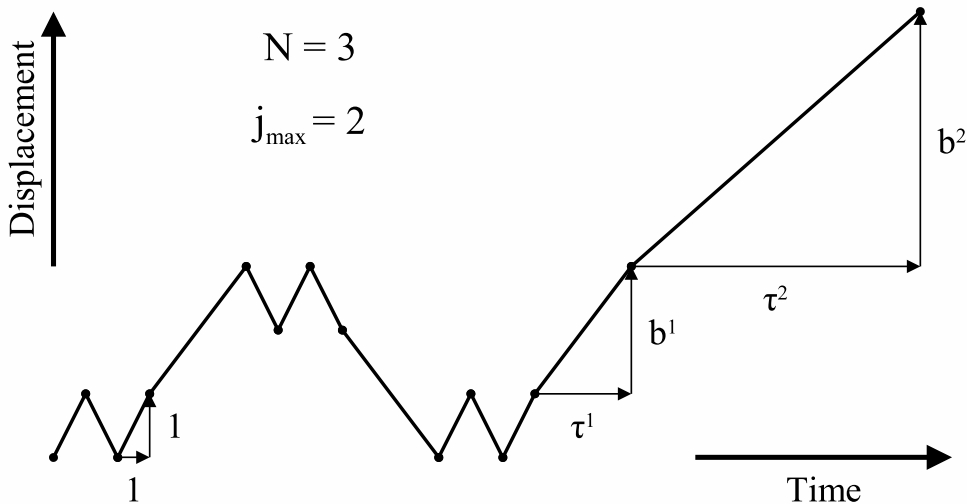}
\caption{Schematic trajectory of hierarchically ordered steps $(\tau ^j,\, b^j)$ presented, for simplicity, for
$N=3$, $j_{MAX}=2$ and calibration parameters $\tau _0=1,\, b_0 =1$. Obviously, in our calculations we assumed 
$j_{MAX}\gg 1$. Herein, the extreme event or black swan is defined by the pair of components 
$(\tau ^{j_{MAX}=2},b^{j_{MAX}=2})$ directed by $j_{MAX}=2$.}
\label{figure:BSwan}
\end{center}
\end{figure}

In Figure \ref{figure:BSwan}, the schematic illustration of the above considerations is shown by using a part 
of the trajectory or random walk realization consisting of hierarchically ordered steps 
$(\tau _0\, \tau ^j,b_0\, b^j)$, for $j=0,1,2$.
Herein, we neglected (i) fluctuation of the number of hierarchy levels $j$s as well as (ii) their random 
succession. Thus we plotted the ordered trajectory within the time-space frame of coordinates. In fact, we made 
a transformation from the stochastic hierarchy to its deterministic counterpart.
%in the manner which keeps fractal (self-similarity) dimensions of both trajectories the same. 
The explanation of the concept of black swans becomes now more convenient.

We can easily derive the useful relation between the single-step mean-square displacement $\left<x^2\right>$ of the 
simulated trajectory and the maximal level, $j_{MAX}$, of the hierarchy contained in it. Indeed, the level $j_{MAX}$ 
defines the extreme event or black swan by the pair of components $(\tau _0\, \tau ^{j_{MAX}},b_0\, b^{j_{MAX}})$. 
This quantity, for a large number of steps $L\gg 1$ or $j_{MAX}\gg 1$, is given by
\begin{eqnarray}
\left<x^2\right>\approx &(b_0)^2\, \left(\frac{N^{j_{MAX}}}{L}\, (b^2)^0+\frac{N^{j_{MAX}-1}}{L}\, (b^2)^1+
\frac{N^{j_{MAX}-2}}{L}\, (b^2)^2+\ldots +\frac{N^0}{L}\, (b^2)^{j_{MAX}}\right)
\label{rown:x2}
\end{eqnarray}
where now $\left<\ldots \right>$ means an average over the random walk steps. That is, we used here a kind of an 
ergodic hypothesis. 

From Eqs. (\ref{rown:X2}) and (\ref{rown:x2}) we obtain
\begin{eqnarray}
\left<X^2\right>(L)
%&\approx &(b_0)^2\, \left(N^{j_{MAX}}\, (b^2)^0+N^{j_{MAX}-1}\, (b^2)^1+
%N^{j_{MAX}-2}\, (b^2)^2+\ldots +N^0\, (b^2)^{j_{MAX}}\right) \nonumber \\   
&\approx & (b_0)^2\, N^{j_{MAX}}\, \frac{(b^2/N)^{j_{MAX}+1}-1}{b^2/N-1}\approx (b_0)^2\, \left\{
\begin{array}{cc}
\frac{1}{1-N/b^2}\, (b^2)^{j_{MAX}}, & \mbox{for $b^2/N >1$} \\
\frac{1}{1-b^2/N}\, N^{j_{MAX}}, & \mbox{for $b^2/N <1$} 
\label{rown:displace}
\end{array}
\right.
\end{eqnarray}
or
\begin{eqnarray}
\left<X^2\right>(L)&\approx & (b_0)^2 \, \left\{
\begin{array}{cc}
\frac{\left(1-1/N\right)^{2/\beta }}{1-1/N^{2/\beta -1}}\, L^{2/\beta }, 
& \mbox{for $\beta < 2$} \\ 
\frac{1-1/N}{1-b^2/N}\, L, & \mbox{for $\beta > 2$} 
\label{rown:displaceX}
\end{array}
\right.
\end{eqnarray}
as well as
\begin{eqnarray}
\left<X^2\right>(L)&\approx & \left\{
\begin{array}{cc}
\frac{1}{1-1/N^{2/\beta -1}}\, (x_{MAX})^2, 
& \mbox{for $\beta < 2$} \\ 
\frac{1-1/N}{1-N^{2/\beta -1}}\, L, & \mbox{for $\beta > 2$} 
\label{rown:displaceXX}
\end{array}
\right.
\end{eqnarray}
where the number of steps 
\begin{eqnarray}
L\approx N^{j_{MAX}}+N^{j_{MAX}-1}+N^{j_{MAX}-2}+\ldots +N^0 \approx \frac{1}{1-1/N}\, N^{j_{MAX}} 
\label{rown:time}
\end{eqnarray}
and the single-step displacement of black swan is defined by 
\begin{eqnarray}
x_{MAX}= b_0\, b^{j_{MAX}}. 
\label{rown:xMAX}
\end{eqnarray}
Herein, the marginal case of $\beta =2$ is not considered. Apparently, for $\beta <2$, i.e. if $\left<X^2\right>$ 
scales with $L$ according to some power-law, $\left<X^2\right>$ is fully determined by $x_{MAX}^2$. This is exactly 
what we need for illustration of our considerations. That is, the quantities (herein, the mean-square displacement 
of the process) which characterize the system are mainly expressed by the corresponding ones (herein, a single-step
displacement) which define the black swans.

Note that in the case of the WM-CTRW we have results analogous to 
those presented by Equations (\ref{rown:displaceX}) and (\ref{rown:displaceXX}). However, their derivation is much 
more complicated in this case. The threshold property exhibited by these equations is typical of the behavior of other 
key quantities, like volatilities or correlation functions.

As expected, for Brownian motions (that is for the case of $\beta >2$ in Equations (\ref{rown:displace}), 
(\ref{rown:displaceX}) and (\ref{rown:displaceXX})), the factor preceding $L$ in the second formula in 
(\ref{rown:displaceX}) equals, in fact, to the corresponding factor present in Formula (\ref{rown:x22}). Here, the 
absence of factor $2$ is only caused by the absence of fluctuations of interevent times.

Now, we are ready to answer the question concerning the distribution of the single-step displacements 
$\mid x\mid =b_0\, b^j$ of the walker. This answer is based on the change of variables from $j$ to $\mid x\mid $. 
Hence, the corresponding (normalized) distribution $\tilde{w}(\mid x\mid )$ takes the Pareto form
\begin{eqnarray}
\tilde{w}(\mid x\mid )\approx \frac{1}{b_0}\, \frac{\beta }{(\mid x\mid /b_0)^{\beta +1}},
\label{rown:wx}
\end{eqnarray}
where $\mid x\mid \ge b_0$. In fact, Equation (\ref{rown:wx}) holds only for $\mid x\mid \gg b_0$. It is well known 
\cite{MDS,KK} that the power-law distribution of events leads to the Fr\'echet distribution of extreme events. For 
asymptotic values of the argument, this distribution preserves the power-law of exponent $\beta + 1$ with the 
power-law correction to scaling of exponent $\beta $. Note that Equation (\ref{rown:wx}) is valid for all values of 
$\beta $. That is, black swans are always present in the system. However, only for $\beta <2$ their influence 
dominates.
   
To conclude this discussion, we can say that the linear size of the walk (herein, the mean-square displacement)
is controlled by extreme events or black swans if this size scales with the number of steps according to some 
power-law, i.e. if the walk is a kind of fractional random walk. Otherwise, the Brownian motions dominate and no 
influence of black swans is observed (although they are present in the system). That way the threshold functions 
as a discriminator of black swans. 

It is evident now, that we can identify an event as a sustained dragon-king if $j_d$ index of this event distinctly 
exceeds that of $j_{MAX}$.  

\subsection{Comparison of theoretical predictions with simulations}\label{section:comparis}

Herein, we restrict our simulations to an important example of a basic random walk given by a fractional Brownian 
motion. That is, we study a confined region of a superdiffusion phase defined by $1/\beta $ only slightly smaller 
than $1/\alpha $. Note that inequality $1/\beta <1/\alpha $ is equivalent to $v<1$, i.e. it corresponds to the case 
where the velocity of the walker for the higher hierarchy level is smaller. In this case, each moment of the 
arbitrary non-negative order of the (multi-step) displacement is finite for finite times \cite{KS,KutS}. This choice 
of such moments arises from empirical evidences that these moments are always finite. 
Other choices, concerning other diffusion phases, would also be worth studying.
The dragon-king is located "manually" inside the simulated time series in such a way that inequality 
$\Delta t_{MAX}\ll t_L,\, t_R$ is obeyed.

\subsubsection{Results for sustained dragon-king}

Now, we discuss the case where the multiplicative factor preceding $v_d^2$ in Equation (\ref{rown:CLCRC}) 
is positive, which is easy to fulfil. That is, we consider inequality 
$\frac{\Delta t}{t_{tot}}<\gamma _d\left(1-\frac{\gamma _d}{1-\Delta t/t_{tot}}\right)$. This inequality
is only slightly stronger than $\frac{\Delta t}{t_d}<1$, yet needed for the derivation of any of our expression 
for $C_d(\Delta t)$ in the case of the appearance of the sustained dragon-king.

In cumulative Table \ref{table:joty} we present data (in a form of four inverted pyramids of numbers), which 
define unnormalized statistics, $S(j)$, of hierarchy levels $j$s (cf. Section \ref{section:Wsda}). These data were 
obtained for fixed common parameters $\tau _0=1.0,\, \tau =2.520,\, v_0=1.0, v=0.992$, and $N=4$. They were 
used to prepare four trajectories within the WM-CTRW formalism in the presence of sustained dragon-kings. For 
instance,
the row numbered by level $j=3$ gives, at intersection with the second column, the number which says how many times 
(herein, it is $809231$) this level appeared in the first trajectory. This trajectory contains the only sustained 
dragon-king defined by index $j_d=13$. This index is shown in Table \ref{table:joty} by bold number $1$, at the 
intersection of row numbered by level $j=13$ with the second column, again. The successive columns from three to 
five contain analogous unnormalized statistics but for an increasing $j_d$ values\footnote{Equations 
(\ref{rown:sustaineddk}) and (\ref{rown:dynstatbj}) precisely define the role of index $j_d$ used here.}, i.e. 
$j_d=15, 17$, and $19$; that is, for increasing sustained dragon-kings.  
\begin{table}[htb]
\begin{center}
\caption{Four unnormalized statistics $S(j)$ of hierarchy levels $j$s obtained for four sustained dragon-kings.}
\label{table:joty}
\begin{tabular}{|c||r|r|r|r|}
\hline 
Level $j$ & $S(j)$ for $j_d=13$ & $S(j)$ for $j_d=15$ & $S(j)$ for $j_d=17$ & $S(j)$ for $j_d=19$ \\
\hline 
\hline 
$0$ & $51763445$ & $51439530$ & $49410801$ & $36182538$ \\
\hline 
$1$ & $12948042$ & $12866869$ & $12360063$ & $9049915$ \\
\hline
$2$ & $3234819$ & $3214452$ & $3087567$ & $2260332$ \\
\hline
$3$ & $809231$ & $804047$ & $772246$ & $565401$ \\
\hline 
$4$ & $202591$ & $201289$ & $193303$ & $141499$ \\
\hline
$5$ & $50583$ & $50521$ & $48326$ & $35374$ \\
\hline
$6$ & $12773$ & $12704$ & $12211$ & $8895$ \\
\hline
$7$ & $3162$ & $3141$ & $3012$ & $2212$ \\
\hline
$8$ & $811$ & $801$ & $765$ & $565$ \\
\hline
$9$ & $192$ & $191$ & $181$ & $128$ \\
\hline
$10$ & $48$ & $47$ & $43$ & $29$ \\
\hline
$11$ & $6$ & $6$ & $6$ & $4$ \\
\hline
$12$ & $5$ & $5$ & $4$ & $3$ \\
\hline
$13$ & ${\bf 1}$ & $0$ & $0$ & $0$ \\
\hline
$14$ & $0$ & $0$  & $0$ & $0$ \\
\hline
$15$ & $0$ & ${\bf 1}$ & $0$ & $0$ \\
\hline
$16$ & $0$ & $0$ & $0$   & $0$ \\
\hline
$17$ & $0$ & $0$ & ${\bf 1}$ & $0$ \\
\hline
$18$ & $0$ & $0$ & $0$ & $0$ \\
\hline
$19$ & $0$ & $0$ & $0$ & ${\bf 1}$ \\
\hline
\end{tabular}
\end{center}
\end{table}
The common bottom of all hierarchies (represented in Table \ref{table:joty} by inverted pyramides of numbers) is 
placed at the level $j=j_{MAX}=12$, that is above any $13\le j=j_d\le 19$. 

So, all sustained dragon-kings are 
marked in Table \ref{table:joty} by the bold number $1$. They are placed from the second to fifth column at  
intersections with the corresponding rows indexed by levels from $j=j_d=13$ to $j=j_d=19$. That is, these levels 
systematically move away from the common bottom of hierarchies.  

Importantly, the left part of trajectory (preceding the dragon-king appearance) is common for all dragon-kings. 
The right border of this part is fixed defining the beginning of a dragon-king. Because the total duration time, 
$t_{tot}$, of all trajectories is the same, the duration time, $t_R$, of the part of trajectory placed on the 
right-hand side of the dragon-king decreases as the duration time $t_d$ increases (i.e. when the index $j_d$ 
increases). Therefore, the 
statistics $S(j)$ of hierarchy levels $j$s, shown in Table \ref{table:joty}, decreases as the level defining 
the sustained dragon-king, $j_d$, is rised (because then the corresponding trajectory or time series, placed on the 
right-hand side of the 
dragon-king, is shorter). Therefore, for example, the number $128$ placed at intersection of the fifth column and the 
row denoted by the index level $j=9$ is distinctly smaller than the number $192$ placed at the same row but at the 
intersection with the second column. We hope that Table \ref{table:joty} well illustrates the hierarchical structure 
of any (long) trajectory simulated within the WM-CTRW formalism. Moreover, the corresponding localisations of the 
sustained dragon-kings in the space of hierarchy levels relative to the inverted pyramids are also well delineated. 

\begin{figure}[htb]
\begin{center}
\includegraphics[width=115mm,angle=270,clip]{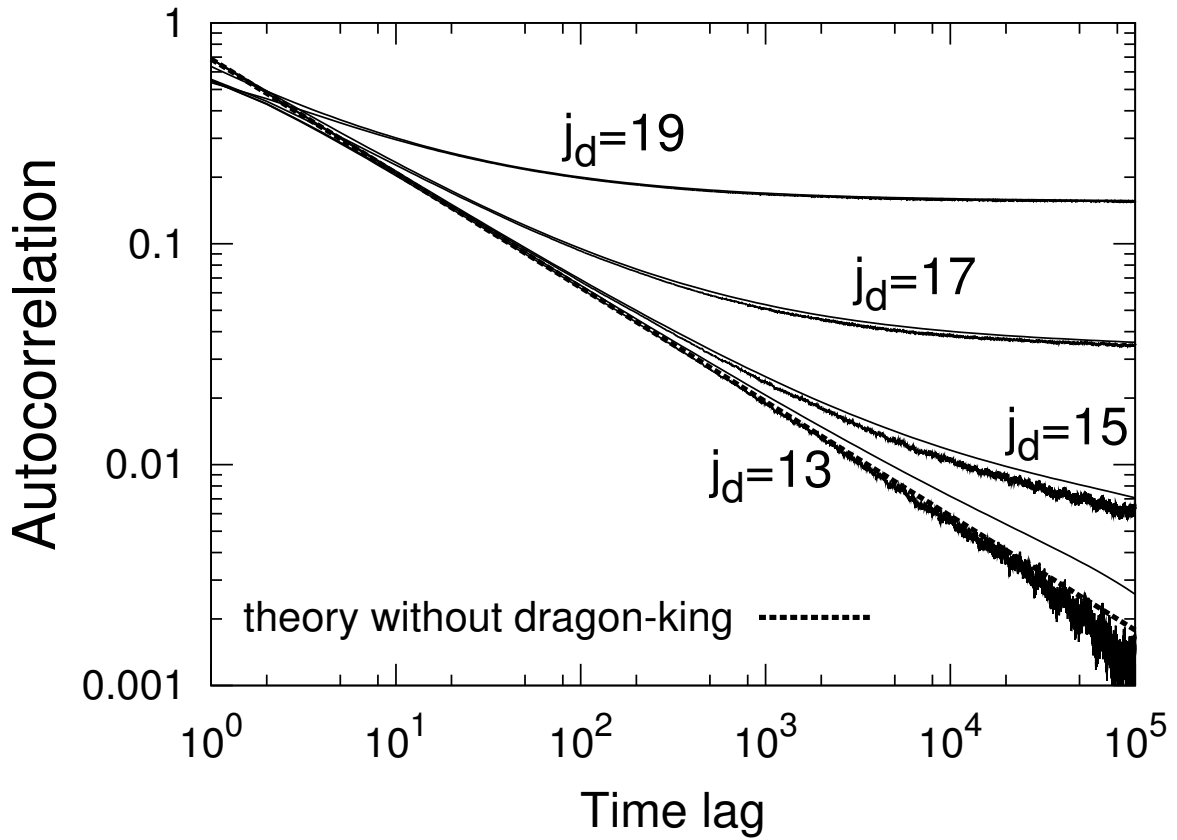}
\caption{Comparison of the prediction of Equation (\ref{rown:CLCRC}) (solid thin curves) with results of simulations 
(dispersed solid thick curves) for four different values of $t_d/\Delta t_{MAX}=1.653,\, 10.496,\, 66.651$, and 
$423.263$ which correspond to $j_d=13,\, 15,\, 17$, and $19$, respectively. The dashed curve represents the 
prediction of Equation (\ref{rown:DXtst}), i.e. the prediction for the time series in the absence of a dragon-king. All 
curves were calculated for the same values of $\tau =2.52,\; v=0.992$, and $N=4$. Notably, theoretical predictions 
for $j_d=17$ and $19$ are almost indistinguishable (within the resolution of the plot) from results of the 
corresponding simulations for the whole range of time lag $\Delta t$. The presence of the sustained dragon-king twists 
upward the curve deviating it from the straight line (in the $\log-\log$ plot).}
\label{figure:drag_c}
\end{center}
\end{figure}

In Figure \ref{figure:drag_c}, we compare the prediction of Formula (\ref{rown:CLCRC}) (thin solid curves) 
with results of simulation (dispersed thick solid curves) for four different values of $t_d$, namely, 
$t_d/\Delta t_{MAX}=1.653,\, 10.496,\, 66.651$, and $423.263$, which correspond to $j_d=13,\, 15,\, 17$, and $19$, 
respectively. Note that the current width of the time-window, $\Delta t$, (called also {\it time lag}) ranges from 
$\Delta t =dt$ up to $\Delta t=\Delta t_{MAX}=10^5\, dt$ with the time step, $dt=1$, while 
$t_{tot}=1400\,\Delta t_{MAX}$ is the same for all statistics $S(j)$ shown in Table \ref{table:joty}.
The sustained dragon-kings' time lags were calculated for the same value of $\tau =2.520$ and the 
single-step displacements were calculated for the common $b=v\, \tau =2.50$ (where $b_0=v_0\tau _0=1.0$). 

The upward convexity of the curves in Figure \ref{figure:drag_c} is due to the presence of the corresponding sustained 
dragon-king. As we expected, the agreement shown in Figure \ref{figure:drag_c} between the prediction of Formula 
(\ref{rown:CLCRC}) and data obtained from simulations becomes better the further the dragon-king is located from the 
top of the hierarchy (see the location of bold number $1$ in Table \ref{table:joty}). The best agreement is obtained 
for the largest $j_d=19$. In other words, the dragon-kings defined by $j_d$ smaller than $19$ slightly positively 
deviate their corresponding velocity autocorrelation functions, $C_d(\Delta t)$, from their simulational counterparts. 

The quantity $C(\Delta t)$, controlled by black swans and given by Expression (\ref{rown:DXtst}), is also 
plotted in Figure \ref{figure:drag_c} (the dashed curve) as a reference VAF, i.e. this VAF was calculated for the 
absence of a dragon-king.  
 
\subsubsection{Results for the shock dragon-king}

\begin{figure}[htb]
\begin{center}
\includegraphics[width=130mm,angle=270,clip]{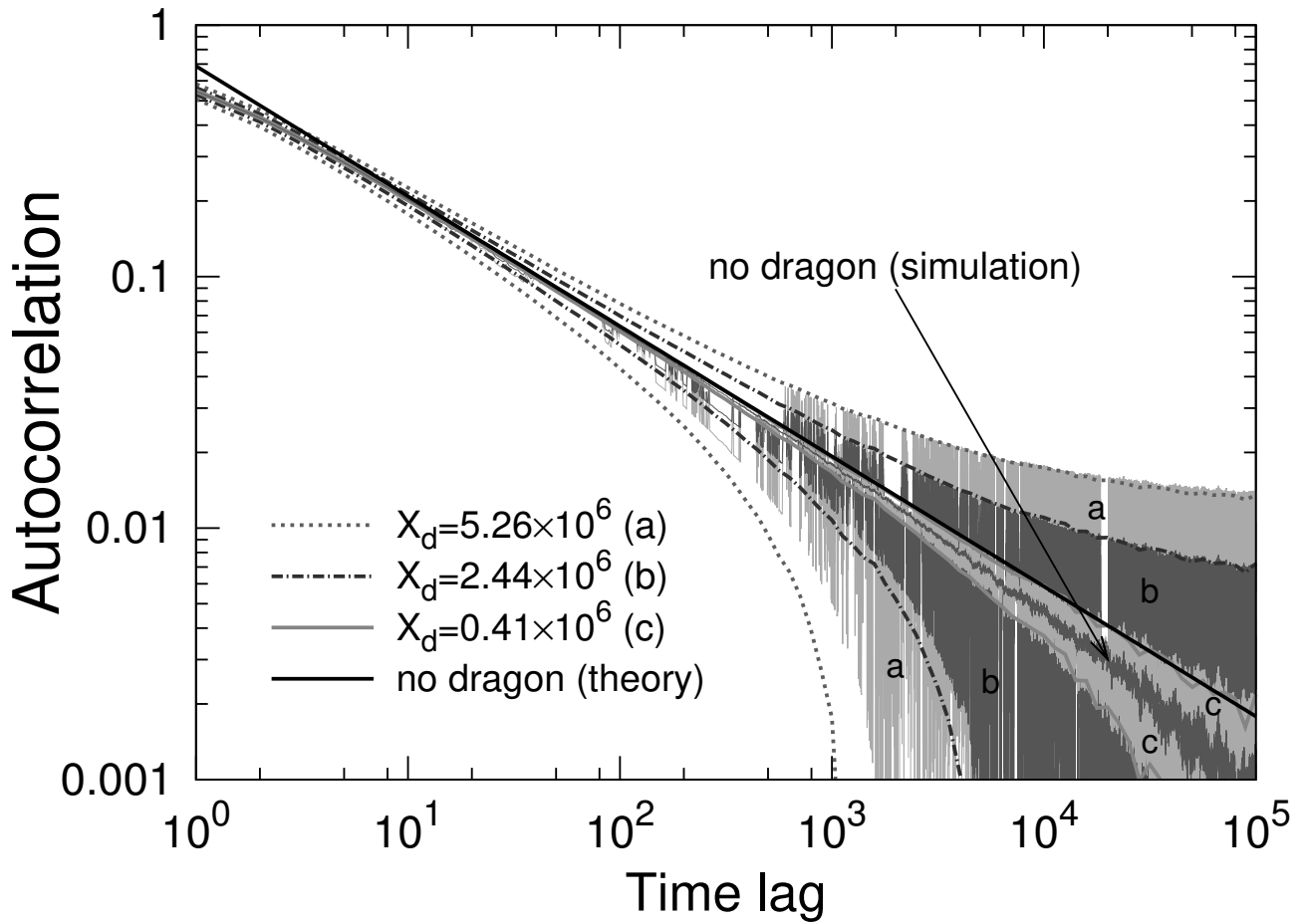}
\caption{Comparison of the prediction of Formula (\ref{rown:envelop}) (dotted and  dashed-dotted curves as well as grey 
solid one) 
with results of simulations (corresponding regions having different greyness, additionally marked by a, b and c) for 
three different values of $X_d=5.26,\, 2.44$, and $0.41\, [\times 10^6]$. The black solid curve shows the prediction 
of Formula (\ref{rown:DXtst}), i.e. the prediction for the time series simulated in the absence of a shock 
dragon-king; the corresponding simulated VAF is given for this time series by the innermost (dark) region. All curves 
were obtained for values of $\tau =2.52,\; v=0.992$, and $N=4$, the same as for the case of the sustained dragon-king.} 
\label{figure:shock_c}
\end{center}
\end{figure}

In Figure \ref{figure:shock_c}, we present results of simulations of the VAF for three different 
values of the shock size $X_d=0.41, 2.44$ and $5.26\, [\times 10^6]$ which corresponds to $X_d/x_{MAX}=1.90,\, 11.3$, 
and $24.4$, respectively, where $x_{MAX}$ is the maximal spatial value of the random walk's single step belonging to 
the simulated hierarchical WM-CTRW trajectory in the absence of a dragon-king. 
Remarkably, all these shocks are fully exogenous as they were taken from outside of the spatio-temporal structure of
time series or random walks. The simulated trajectories have also the total time $t_{tot}=1400\,\Delta t_{MAX}$ and, 
except for the presence of the dragon-king, all trajectories are identical. 

A striking property of simulated VAFs is their dispersion behaving like a certain instability. Therefore, it is 
more convenient to use a formula that only describes the dispersion of the data. In principle, such a formula could 
be obtained by replacing the sum of velocities $v_L+v_R$ in Formula (\ref{rown:dragonshoc}) by its dispersion 
$\sigma =\sqrt{\left<(v_L+v_R)^2\right>}=\sqrt{2}\sqrt{\sigma _v^2+C(2\Delta t)}$, where 
$\sigma _v^2=\left<v_L^2\right>=\left<v_R^2\right>$. This is allowed because we assumed, for simplicity, that a shock 
does not change the type of the random walk. 

Moreover, our approach allows us to study a more realistic case, e.g. in which dispersion is smaller 
than $\sigma $ defined above. Hence, we propose a more flexible stationary formula
\begin{eqnarray}
C_d(\Delta t)=C(\Delta t)\pm \sqrt{2}\, \sigma _f\,\frac{X_d}{t_{tot}-\Delta t}
-\left(\frac{X_d}{t_{tot}-\Delta t}\right)^2,
\label{rown:envelop}
\end{eqnarray}
where $\sigma _f\stackrel{\rm def.}{=}\sqrt{f\sigma _v^2+C(2\Delta t)}$ and the phenomenological factor or weight,
$0<f\leq 1$, is the same for all trajectories. As it is seen, also in the case of the shock dragon-king we transformed
the non-stationary Expression (\ref{rown:dragonshoc}) to more useful stationary Expression (\ref{rown:envelop}). 

Comparison of the prediction of Equation (\ref{rown:envelop}) 
with the data obtained by simulation is shown in Figure \ref{figure:shock_c}. In this figure, only small deviations are 
seen for the choice of the factor $f=0.30$. The data scatter is reasonably small but it increases with the increase of the ratio 
$X_d/t_{tot}$. The origin of this scatter comes from fluctuations of the simulated trajectory, unfortunately resulting 
also in a spontaneous artificial trend (e.g. as a deviation from the power-law in the absence of the dragon-king). 
Additionally, this trend can be supported by the finite size of the simulated time series.  

Note that further simplification of Equation (\ref{rown:envelop}) is also possible 
\begin{eqnarray}
C_d(\Delta t)=C(\Delta t)\pm \sqrt{2}\, \sigma _f\, \frac{X_d}{t_{tot}}-\left(\frac{X_d}{t_{tot}}\right)^2,
\label{rown:envsimpl}
\end{eqnarray}
if the strong but reasonable inequality $\Delta t_{MAX}/t_{tot}\ll 1$ is obeyed. 

\section{Summary and concluding remarks}\label{section:sumconclud}

In the present work we discussed the following issues. 
\begin{itemize}
\item[(i)] We considered the influence of two distinctive types of dragon-kings on the velocity autocorrelation 
function, namely the sustained dragon-king and shock dragon-king. 
\item[(ii)] By simulations and by theoretical analysis, we found that the dragon-king influence decisively changes the
original VAFs calculated, for instance, within the hierachical Weierstrass-Mandelbrot Continuous-Time Random Walk 
formalism for a wide range of time intervals. The influence of both types of dragon-kings is 
well pronounced but quite different (cf. the corresponding plots shown in Figures \ref{figure:drag_c} and 
\ref{figure:shock_c}). 
Remarkably, the results obtained by simulations agree well with the corresponding predictions
of our simple theoretical Formulas (\ref{rown:CLCRC}) and (\ref{rown:envelop})  (see again Figures 
\ref{figure:drag_c} and \ref{figure:shock_c}).
\item[(iii)] Furthermore, several intermediate formulas, e.g. (\ref{rown:covd}), (\ref{rown:Cfingen}), or 
(\ref{rown:dragonsh}), derived in this work for $C_d(\Delta t)$, can be applied to more complex cases where, for 
instance, (a) after the dragon-king appearance the random walk is changed, (b) the random walk with drift is considered 
and (c) dragon-kings cluster in the system. 
\end{itemize}

As it is apparent from Figure \ref{figure:drag_c}, the presence of the sustained dragon-king throws the system 
away from the state controlled by black swans.  
That is, the 'finger-print' of this dragon-king is significant because it convexly twists upward the autocorrelation 
curve dependence on the time lag and distinctly deviates it positively from the power-law generated by black swans 
(cf. Section 
\ref{section:stMW-CTRW} for details). This deviation is one of the main features which distinguishes $C_d(\Delta t)$ 
from usual $C(\Delta t)$. This difference allows one to distinguish a power-law relaxation, controlled by black 
swans, from the decay controlled by the sustained dragon-king. Moreover, this difference can be so distinct that it 
can effectively be applied as a tool to detect the sustained dragon-king from an empirical time series.  

The plot in Figure \ref{figure:shock_c} shows that the shock dragon-king also significantly changes $C(\Delta t)$. 
However, this change
is different than the change caused by the sustained dragon-king. This change is also well seen by direct comparison 
of Formulas (\ref{rown:CLCRC}) and (\ref{rown:envelop}). Noticeably, the scatter of the data  
shown in Fig. \ref{figure:shock_c} indicates some instability of the system after the appearance of the shock 
dragon-king. Indeed, if the empirical VAF reveals such an instability then we can anticipate that the corresponding 
empirical time series contains the shock dragon-king. That is, such an anomalous VAF indicates shocks.

\begin{acknowledgments}
We thank Armin Bunde for stimulating discussion. This work was partially supported by the Grant No. 119 
awarded within the First Competition of the Committee of Economic Research, organized by the National Bank of 
Poland.
\end{acknowledgments}

\end{document}